\documentclass[12pt]{article}
\usepackage{geometry}
\usepackage[T1]{fontenc}
\usepackage{layaureo}
\usepackage{graphicx}
\usepackage{tabularx}
\usepackage{booktabs}
\usepackage{siunitx}
\usepackage{amsmath}
\usepackage{amssymb}

\begin{document}
\title{\textbf{tvf-EMD based time series analysis of $^{7}$Be of the CTBTO-IMS network}}
\author{A. Longo$^a$, S. Bianchi$^a$, W. Plastino$^a$}
\date{\small $^a$\textit{Department of Mathematics and Physics, Roma Tre University, Via della Vasca Navale, 00146, Rome, Italy}}

\maketitle
\begin{abstract}
A methodology of adaptive time series analysis based on Empirical Mode Decomposition (EMD) has been employed to investigate $^{7}$Be activity concentration variability, along with temperature. Analysed data were sampled at ground level by 28 different stations of the CTBTO-IMS network. The adaptive nature of the EMD algorithm allows it to deal with data that are both nonlinear and non-stationary, making no a priori assumptions on the expansion basis.
Main purpose of the adopted methodology is to characterise the possible presence of a trend, occurrence of AM-FM modulation of relevant oscillatory modes, residuals distributions and outlier occurrence. Trend component is first estimated via simple EMD and removed. The recent time varying filter EMD (tvf-EMD) technique is then employed to extract local narrow band oscillatory modes from the data. To establish their relevance, a denoising step is then carried out, employing both the Hurst exponent as a thresholding parameter and further testing their statistical significance against white noise. The ones that pass the denoising step are considered to be meaningful oscillatory modes of the data, and their AM-FM modulation is investigated.
Possible applications of the adopted methodology regarding site characterisation and suggestions for further research are given in the conclusions.
\end{abstract}
\section*{Introduction}
Time series analysis has relevant applications in many fields of science, particularly geophysics, e.g. in seismology, radionuclide concentration monitoring and tracing, climate and atmospheric physics, oceanography, geomagnetism \cite{Bizzarri_2010,Plastino_2010,Lee_2015,Blender_2011,Alford_2016,Wunsch_2007,Balasis_2012,Occhipinti_2011}. It is also applied in site characterisation studies in order to investigate possible coupling of environmental noises to detector instrumentation. Relevant examples in this regard are seismic and Newtonian Noise (NN) \cite{Driggers_2012,Fiorucci_2018}. Beside applications in Physics, time series analysis studies also has relevant impact, among others, in
wind speed forecasting \cite{Niu_2001}, physiology \cite{Soehle_2008}, speech pattern recognition \cite{Manjula_2012}, power quality events classification \cite{Koh_2012}. In this paper a methodology of adaptive time series analysis based on Empirical Mode Decomposition (EMD) \cite{Huang_1998,Huang_book,Wu_2004} is applied to $^{7}$Be activity concentration data, sampled at ground level, to investigate their variability. Aim of the analysis, is to characterise time series of data, $x(n)$, in terms of three main components, namely the trend, harmonics and residuals
\begin{equation}
x(n) = t(n) + s(n) + r(n)
\label{eq:time_series}
\end{equation}
where $x(n)$, $t(n)$, $s(n)$, and $r(n)$ stands respectively for time series, trend, extracted signal (harmonics), and noise (residuals), and $n = 1\dots N$, with $N$ the length of the time series. EMD allows to characterise both nonlinear and non-stationary data \cite{Huang_1998}.
The paper is organised as follows.
In Section \ref{methodology} the analysed dataset is briefly described and the adopted methodology is introduced in details. In Section \ref{results}, data analysis results are presented.
\section{Dataset and Methodology}\label{methodology}
All the data analysed in this paper were collected on a daily basis by 28 stations of the International Monitoring System (IMS), a worldwide distributed network set up and maintained by the Comprehensive Nuclear Test Ban Treaty Organisation (CTBTO), which goal is to monitor over CTBT compliance. 
Time series of $^{7}$Be and surface temperature have been selected based on data availability.
The length of a given time series depends on when the corresponding station became operative, and the starting year of the time series varies from 2003 to 2009, while all series end on March 2016.
A more detailed description of the analysed $^{7}$Be dataset can be found in \cite{Bianchi_be}.
In Table \ref{tab:stations_names} the names and the exact location of the analysed stations are listed.
\begin{table}[phtb]\centering
\begin{tabular}{clrrp{0.5\columnwidth}}
\toprule
ID & Location & Latitude [$^\circ N$] & Longitude [$^\circ E$] \\ 
\midrule
RN46 & Chatham Island, New Zealand & -43.82 & 176.48 \\
RN04 & Melbourne, VIC, Australia & -37.73 & 145.10 \\
RN68 & Tristan da Cunha, United Kingdom & -37.07 & -12.31 \\
RN47 & Kaitaia, New Zealand & -35.07 & 173.29 \\
RN01 & Buenos Aires, Argentina & -34.54 & -58.47 \\
RN10 & Perth, WA, Australia & -31.93 & 115.98 \\
RN23 & Rarotonga, Cook Islands & -21.20 & -159.81 \\
RN06 & Townsville, QLD, Australia & -19.25 & 146.77 \\
RN26 & Nadi, Fiji & -17.76 & 177.45 \\
RN09 & Darwin, NT, Australia & -12.43 & 130.89 \\
RN08 & Cocos Islands, Australia & -12.19 & 96.83 \\
RN64 & Dar Es Salaam, Tanzania & -6.78 & 39.20 \\
RN50 & Panama City, Panama & 8.98 & -79.53 \\
RN43 & Nouakchott, Mauritania & 18.14 & -15.92 \\
RN79 & Oahu, Hawaii, USA & 21.52 & -157.99 \\
RN37 & Okinawa, Japan & 26.50 & 127.90 \\
RN72 & Melbourne, FL, USA & 28.10 & -80.65 \\
RN74 & Ashland, KS, USA & 37.17 & -99.77 \\
RN75 & Charlottesville, VA, USA & 38.00 & -78.40 \\
RN17 & St. John's, N.L., Canada & 47.59 & -52.74 \\
RN45 & Ulaanbaatar, Mongolia & 47.89 & 106.33 \\
RN33 & Schauinsland/Freiburg, Germany & 47.92 & 7.91 \\
RN60 & Petropavlovsk, Russian Federation & 53.05 & 158.78 \\
RN71 & Sand Point, Alaska, USA & 55.34 & -160.49 \\
RN61 & Dubna, Russian Federation & 56.74 & 37.25 \\
RN63 & Stockholm, Sweden & 59.41 & 17.95 \\
RN16 & Yellowknife, N.W.T., Canada & 62.48 & -114.47 \\
RN76 & Salchaket, Alaska, USA & 64.67 & -147.10 \\
\bottomrule
\end{tabular}
\caption{\small ID code and exact location of CTBTO International Monitoring System stations analysed in this paper.}
\label{tab:stations_names}
\end{table}
The methodology adopted in this paper is intended to be complementary to the one described in \cite{Bianchi_be} (hereafter referred to as MTsA),
and aims at extending the range of analysis to strongly nonlinear and non-stationary time series. Detailed description of MTsA software and its applications can be found in \cite{Bianchi_be,Bianchi_bexe,Bianchi_uranium,Longo_2018}.

Cosmogenic beryllium-7 time series measured at 28 different sites at ground level
are analysed making use of the Empirical Mode Decomposition (EMD). EMD is an adaptive algorithm first introduced by Huang \cite{Huang_1998,Huang_book,Wu_2004}. It allows to deal with nonlinear and non-stationary time series and to extract oscillatory modes, referred to as Intrinsic Mode Functions (IMFs). To be an IMF, such oscillatory functions must respect the the following two conditions:
\begin{itemize}
\item	The number of extrema and zero crossings must be equal or differ at most by one.
\item The mean of the upper and lower envelope must be zero.
\end{itemize}
IMFs are obtained by the EMD algorithm fitting upper and lower extrema contained in the data with cubic splines, and then subtracting the mean of the upper and lower envelopes. This procedure is iterated until the two aforementioned conditions are met, and this iterative process is referred to as sifting. Having obtained the first IMF, it is subtracted from the data and the process is repeated on the remainder of the time series until a last term is obtained, $t(n)$, which represents the adaptive trend or baseline wandering of the data, if present. This way, a given time series $x(n)$, can be represented by the following expansion \cite{Huang_1998}
\begin{equation}
x(n) = \sum_{j=1}^{K} c_{j}(n) + t(n)
\label{eq:imfs}
\end{equation}
where $c_{j}$ is the jth IMF, and $n=1 \dots N$, with $N$ the number of data samples, and $K$ is the number of IMFs that have been extracted by EMD algorithm. Since IMFs are mono-component or narrow band oscillatory modes, Hilbert analysis provides then meaningful Instantaneous Amplitude (IA) and Frequency (IF) of such modes. Combining the two approaches, the analysis as a whole is referred as Hilbert-Huang spectral analysis (HHSA). EMD is a fully data-driven technique, making no a priori assumptions on the basis functions for the expansion. It is complete, i.e. the sum of the IMFs and the trend term equals the input data and the IMFs are almost orthogonal. A significance test to distinguish IMFs possibly due to noise from significant ones has been introduced in \cite{Wu_2004}. Due to the fact that IMFs of white noise obtained by EMD decomposition are normally distributed, spread lines giving IMF significance against white noise can be obtained.

A known drawback of EMD application to noisy dataset is mode mixing, namely an IMF containing oscillations of widely different scales or different IMFs having very similar ones \cite{Huang_1998}. To deal with mode mixing and intermittency related problems, and to improve its frequency resolution, a recent modification of the EMD algorithm was introduced in \cite{Li_2017}, the time varying filter EMD (tvf-EMD). The concept of IMF is replaced by that of local narrow band signal, based on instantaneous bandwidth. To extract local narrow band signals, B-splines are employed as a filter with time varying cut off frequency. 
The main steps of the sifting procedure, based on the tvf-EMD algorithm, can be found in \cite{Li_2017}, algorithm 3. To simplify the notation, in the remaining of this paper the term IMFs refers to the narrow band signals extracted by the tvf-EMD algorithm.

\begin{figure}
\centering
\includegraphics[scale=0.4]{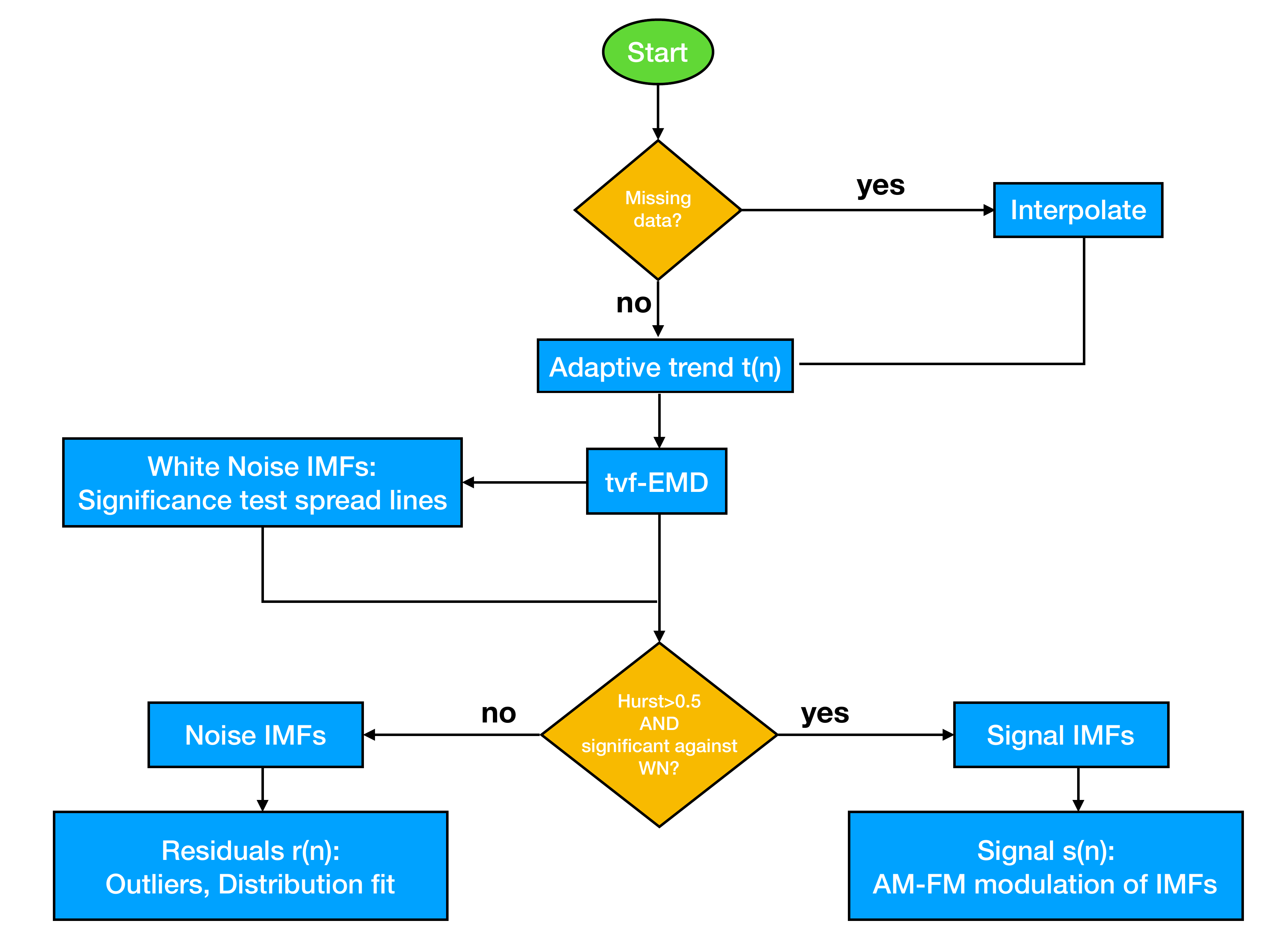}
\caption{Flowchart of the adopted methodology. After removing the trend component $t(n)$, tvf-EMD is performed on both the data and white noise of the same length. IMFs $c_j (n)$ are extracted and their significance is evaluated against white noise and a thresholding parameter based on the Hurst exponent. Summing up the so classified IMFs, noise $r(n)$ and signal $s(n)$ are obtained and separately characterised.}
\label{fig:flowchart}
\end{figure}

The methodology adopted in this paper is based on tvf-EMD and is hereafter described. The flowchart of the methodology of analysis can be found in Figure \ref{fig:flowchart}.
The steps of the adopted methodology are here listed:
\begin{enumerate}
\item Preprocessing: If missing data are present in the data $x(n)$, interpolation is carried out.
\item Detrend: Classical EMD is firstly performed on the data. The last obtained component, $t(n)$, is considered to represent the adaptive trend or baseline wandering of the data, and is removed before further analysis.     
\item Oscillatory modes extraction: To extract local narrow band signals from the data, tvf-EMD algorithm is applied and a set of IMFs, $c_j (n)$, is obtained.
\item Denoising: Following the approach described in \cite{Sundar_2016,Mert_2014}, the IMFs having a Hurst exponent $H < 0.5$ are considered as noise. Following \cite{Wu_2004}, IMFs significance against white noise is also evaluated. Summed together the IMFs satisfying these two criteria are considered to be the residuals $r(n)$. 
\item IMFs that are significant against white noise, and that furthermore have $H > 0.5$, are considered as relevant oscillatory modes of the data. 
\end{enumerate}
While relevant oscillatory modes constituting the signal $s(n)$ can be characterised in terms of AM-FM modulations, obtained residuals $r(n)$ are further characterised in terms of outlier occurrence
following the approach of \cite{Bianchi_be}. Residuals are normalised at zero mean and unit variance, and outliers are defined as those values greater than $3 \sigma$ or less than $-3\sigma$.
Residuals correlations are instead evaluated in terms of their Hurst exponent, as described in \cite{Huang_1998,Rilling_2005}. In order to do so, Hurst exponent of the residuals $r(n)$ is estimated following the approach described in \cite{Rilling_2005} (Equation 5).
At the end of the analysis, performance of the denoising step is evaluated by means of the following parameters, as defined in \cite{Sundar_2016}: 
\begin{itemize}
\item	Mean squared error: $$MSE=\frac{\sum_{n=1}^{N}(x(n)-s(n))^{2}}{N}$$
\item	Mean absolute error: $$MAE=\frac{\sum_{n=1}^{N}|(x(n)-s(n)|}{N}$$
\item	Signal to noise ratio: $$SNR=10\log_{10}\frac{\sum_{n=1}^{N}x(n)^{2}}{\sum_{n=1}^{N}(x(n)-s(n))^{2}}$$
\item	Peak signal to noise ratio: $$PSNR=20\log_{10}\frac{max(x(n))}{RMSE}$$
\item	Crosscorrelation between $x(n)$ and $s(n)$: $$xcorr=\frac{E((s(n)-\mu_{s(n)})(x(n)-\mu_{x(n)}))}{\sigma_{s(n)}\sigma_{x(n)}}$$
\end{itemize}
where $x(n)$ is the input data, $s(n)$ is the extracted signal, RMSE is the square root of MSE and $n=1 \dots N$, with $N$ the number of data samples.
\section{Results and discussion}\label{results}
Results of the analysis are summarised in this section for the 28 stations of the CTBTO IMS network.
\begin{figure}
\centering
\includegraphics[scale=0.75]{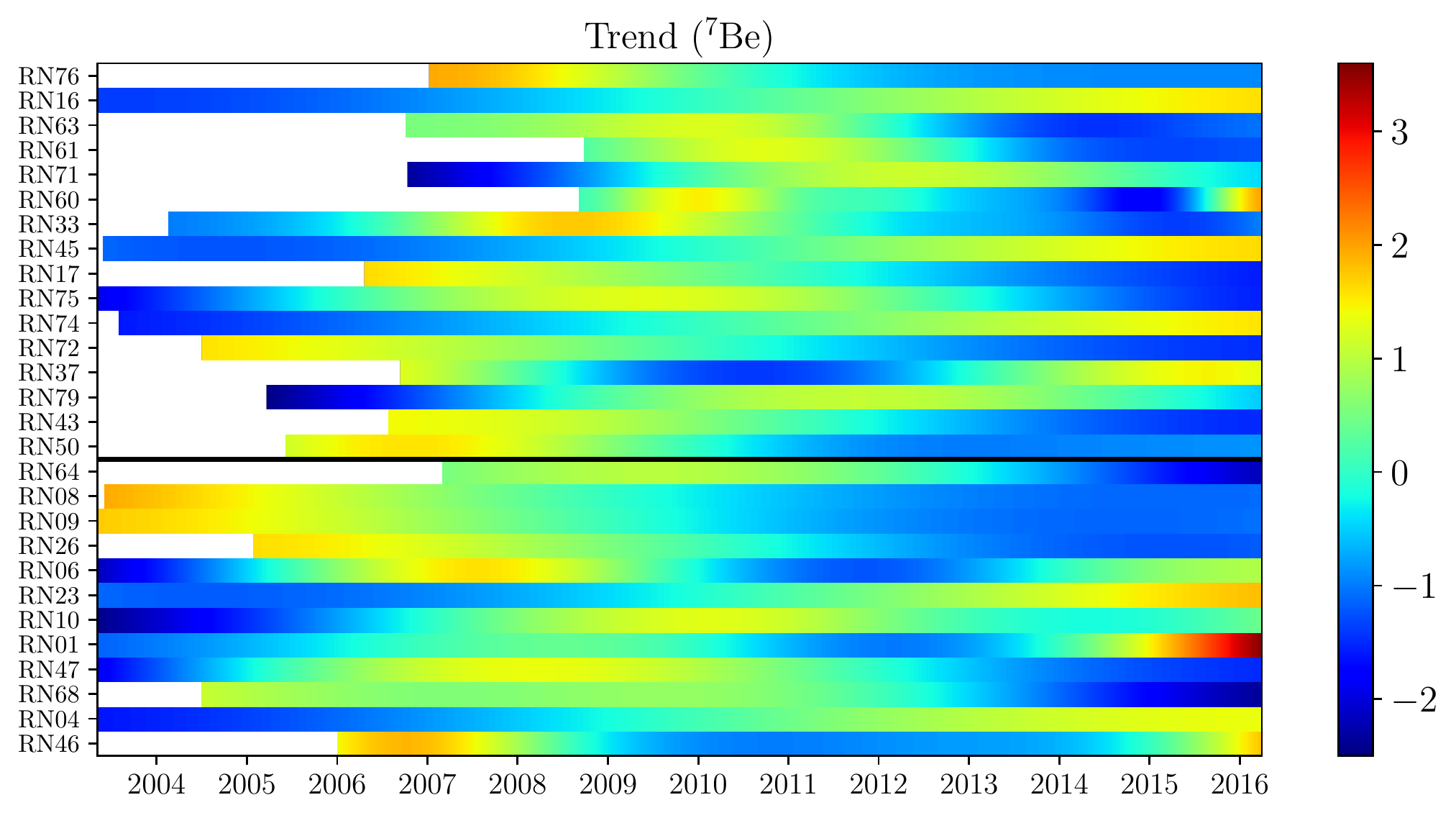}
\caption{$^7$Be adaptive trends as obtained by the last mode of EMD for the 28 stations of the IMS CTBT network. Stations are ordered by latitude and have been normalised to zero mean and unit variance for a meaningful comparison. The horizontal black line divides the Northern hemisphere from the Southern one.}
\label{fig:trends}
\end{figure}
In Figure \ref{fig:trends} the $^{7}$Be extracted adaptive trend are shown. Trends are normalised to zero mean and unit variance for a better comparison. A unique behaviour in terms of latitude cannot be discerned, possibly due to the widely different locations and altitudes of the different stations of the network. Monotonic trends are prevalent in both Northern and Southern hemispheres (divided by the horizontal black line), while there are also non-monotonic trends that express a non-constant growth of beryllium-7 concentrations. It is important to notice that trend behaviour is not completely determined for some stations, since they start years after 2003. $^7$Be trends have been also cross-correlated with the temperature ones. Even though the majority of correlations are high, a clear pattern cannot be evinced.
\begin{figure}
\centering
\includegraphics[scale=0.75]{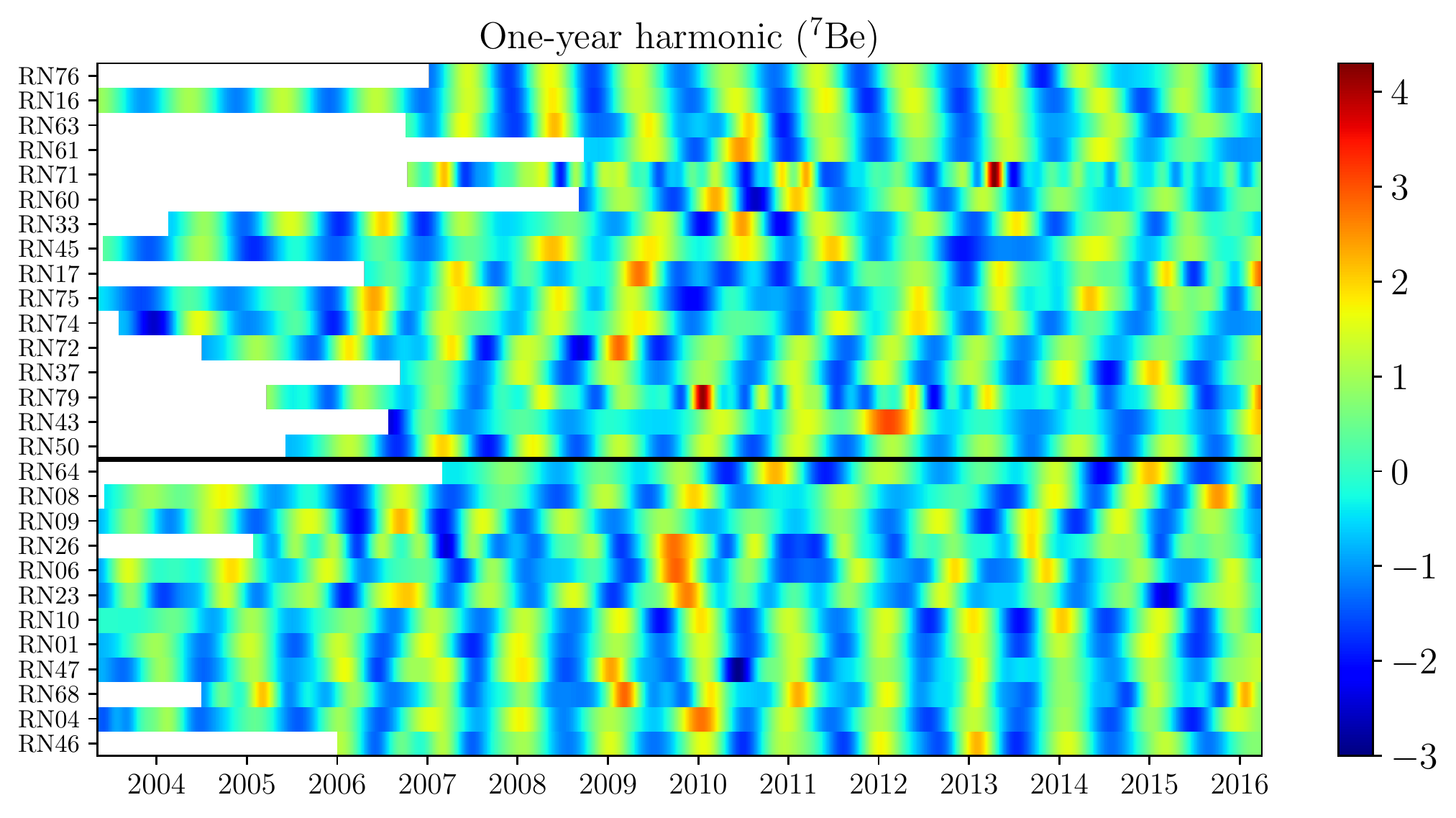}
\caption{$^7$Be annual IMF of the 28 stations of the IMS CTBT network. Stations are ordered by latitude and have been normalised to zero mean and unit variance for a meaningful comparison. The horizontal black line divides the Northern hemisphere from the Southern one.}
\label{fig:annual_be}
\end{figure}
In Figure \ref{fig:annual_be}, the annual IMF of beryllium-7 activity concentration is shown for all the analysed stations. To meaningfully compare the different yearly oscillations, they have been normalised to zero mean and unit variance. Maxima and minima alternate regularly and appear to be shifted in time both in Northern and Southern hemispheres (divided by the horizontal black line). Peaks of the annual oscillation are almost regularly delayed going from the equator to high latitudes, and the same occurs going from the equator to low latitudes. Such behaviour is possibly related to shifting of the Hadley cell and of the inter tropical convergenze zone (ITCZ), as also noted in \cite{Doering_2014}. It should be noted that the considered stations are not uniformly distributed around the globe, and latitudes are not continuous from the top to the bottom of Figure  \ref{fig:annual_be}. Furthermore, stations RN79, RN26, RN06, RN23 and RN04 exhibit the highest value of the yearly peak in 2010. Due to their locations, this is possibly related to the El Ni{\~n}o event occurred in 2009-2010.
\begin{figure}
\centering
\includegraphics[scale=0.75]{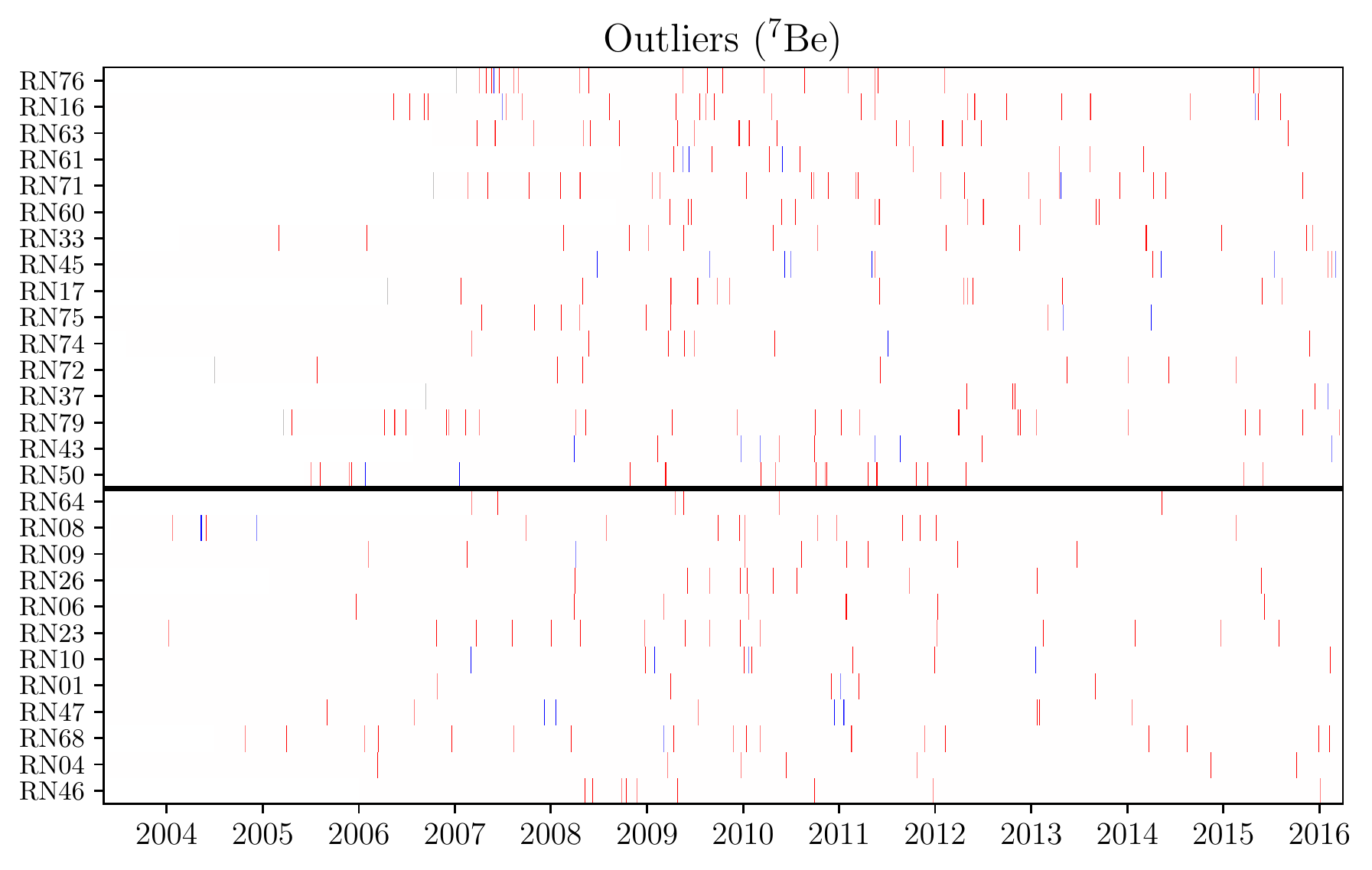}
\caption{Occurrence of outliers higher than $3\sigma$ (red) and lower than $-3\sigma$ (blue) in $^7$Be residuals, for the 28 stations of the IMS CTBT network. Stations are ordered by latitude. The horizontal black line divides the Northern hemisphere from the Southern one.}
\label{fig:res}
\end{figure}
In Figure \ref{fig:res} the occurrence of outliers higher than $3\sigma$ (red) and lower than $-3\sigma$ (blue) in $^7$Be residuals is shown. The two hemispheres are divided by the horizontal black line. A higher number of outliers is observed in the central part of the dataset between 2007 and mid 2012, and in general low values are not frequent, meaning a very rare occurrence of high drops in beryllium-7 concentrations. Furthermore the number of outliers is higher in the Northern hemisphere
compared to the Southern one,
and represents the 61.4\% of the total outliers. Almost 20\% of the Northern hemisphere's outliers are above $4\sigma$, and 4\% is above $5\sigma$. Percentages are lower for the Southern hemisphere, and the number of outliers above $4\sigma$ and $5\sigma$ is half the number of the corresponding Northern outliers.
Residuals correlations have been estimated via the Hurst exponent $H$. All values oh $H$ ranges between $\sim 0.8$ and $\sim 1.0$, indicating strong long-range autocorrelations of residuals time series.
\begin{figure}
\centering
\includegraphics[scale=0.75]{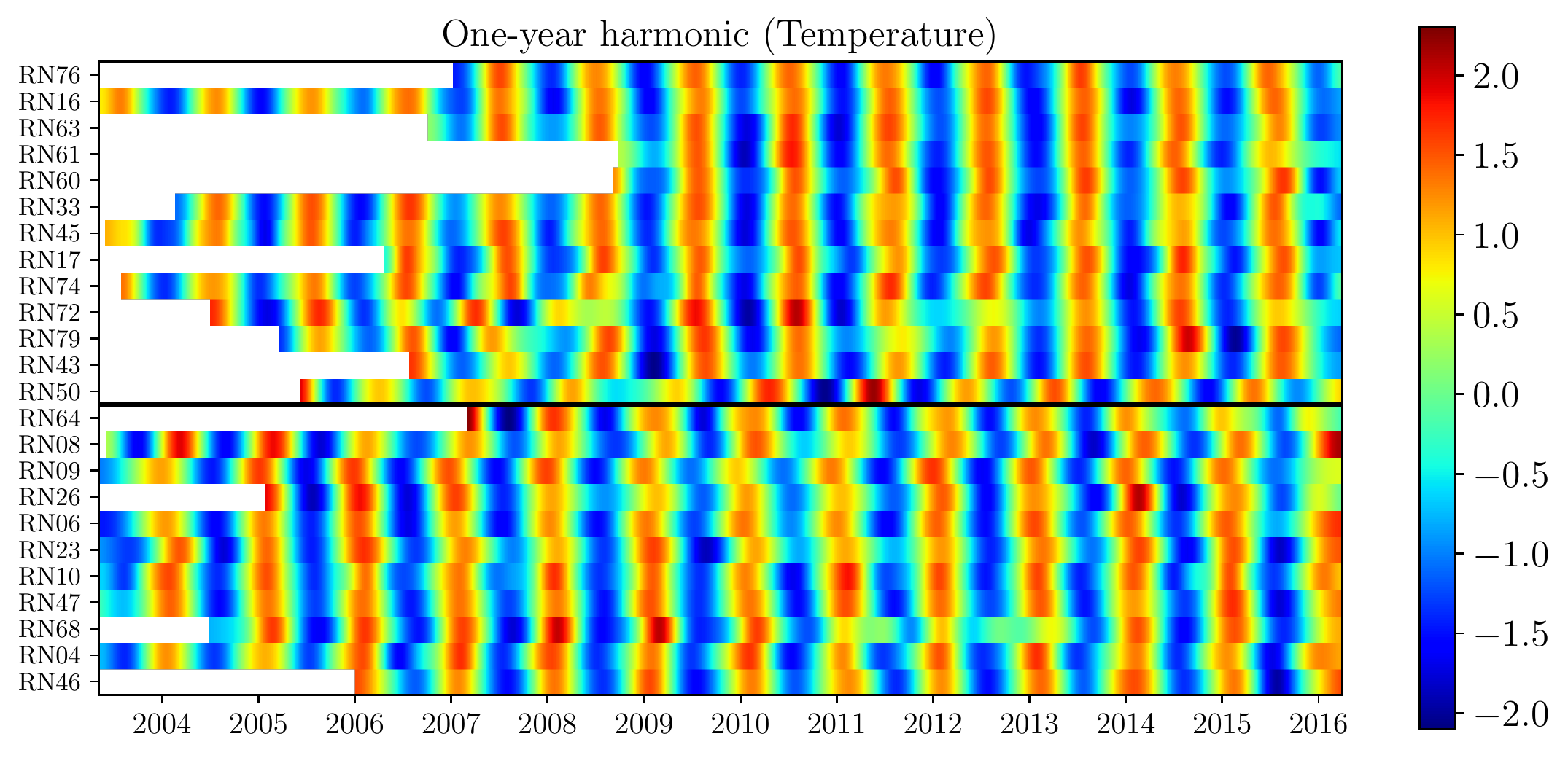}
\caption{Annual IMF of temperature data, normalised to zero mean and unit variance for a better comparison. Stations are ordered by latitude. The horizontal black line divides the Northern hemisphere from the Southern one.}
\label{fig:annual_temp}
\end{figure}
Finally, figure \ref{fig:annual_temp} shows the yearly IMF extracted from temperature data, normalised to zero mean and unit variance for a better comparison. The number of stations is less than 28 since some stations had a poor quality temperature time series. A simple seasonal pattern is observed in this case with a six month shift between Northern and Southern hemisphere maximum temperature, as expected. Cross-correlations between temperature and $^7$Be IMFs have also been evaluated. High correlations have been found in all but two stations, namely RN06 in the Southern hemisphere and RN17 in the Northern hemisphere. These two stations are characterised by a non perfect annual oscillation, as can be seen in Figure \ref{fig:annual_be}.
\section*{Conclusions}
In this paper, data of $^{7}$Be, sampled daily at 28 different stations of the CTBTO IMS and worldwide distributed, have been analysed and characterised using a recent technique based on EMD. Their adaptive trend, yearly variability and outlier occurrence have been investigated.
The annual cycle shows a different pattern at different latitudes, and is delayed at highest and lowest latitudes. Also outliers in the residuals time series occur more frequently in the Northern hemisphere, while autocorrelation properties seem not to depend on site location.

EMD has been proven to be suitable for this analysis, since it is a tool for time-frequency analysis of nonlinear and non-stationary data. Being fully adaptive, it doesn't make any a priori assumption about the basis of expansion, which is carried out in term of amplitude and frequency modulated IMFs. The recently developed tvf-EMD is an extension of such method, improving its frequency resolution while mitigating end effects, mode mixing and intermittency.

Denoising performance has been evaluated selecting a threshold of 0.5 for the Hurst exponent. However, further testing is needed in order to possibly select an optimal Hurst exponent as a thresholding parameter. This will be the object of future research.

\end{document}